\documentclass[12pt]{book}

\usepackage[dvips]{graphicx,color}
\usepackage{makeidx,tsukuba}

\makeauthorindex
\makeindex

\newcommand\paperno{
   \vspace{-8\baselineskip}
   \noindent \underline{\it LANL Report \rm \# LA-UR-03-3323}
   \vspace{6.8\baselineskip}}

\begin{document}

\BookTitle{\itshape The 28th International Cosmic Ray Conference}
\CopyRight{\copyright 2003 by Universal Academy Press, Inc.}
\pagenumbering{arabic}

\chapter{
Evaluation of Production Cross Sections of Li, Be, B in CR
}
\paperno

\author{%
%
%
Igor V.~Moskalenko,$^{1,2}$ Stepan G.~Mashnik$^3$ \\
{\it 
(1) NASA/Goddard Space Flight Center, Code 661, Greenbelt, MD 20771, USA\\
(2) JCA/University of Maryland, Baltimore County, Baltimore, MD 21250, USA\\
(3) Los Alamos National Laboratory, Los Alamos, NM 87545, USA
} \\
}

\section*{Abstract}
Accurate evaluation of production cross sections of light elements
is important for models of CR propagation, galactic chemical
evolution, and cosmological studies. However, experimental
spallation cross section data are scarce and often unavailable to CR
community while semi-empirical systematics are frequently wrong by a
significant factor. We use all available data from
LANL nuclear database together with modern nuclear codes
to produce evaluated production cross sections
of isotopes of Li, Be, B suitable for astrophysical applications.

\section{Introduction}
The accuracy of the nuclear
cross section calculations used in astrophysics is far behind the
accuracy of recent CR measurements
and clearly becomes a factor restraining further progress (see
[4,5] for a discussion).   
Scarce cross section measurements alone can not be used to 
produce a reliable evaluation of the cross sections, while 
current nuclear codes and semi-empirical parametrizations also fall
short of predicting cross section behavior for the whole range 
of target nuclei and incident energies.
We use all available data (including cumulative and isobaric
reactions, and reactions on natural samples of elements)
together with modern nuclear codes
to produce evaluated production cross sections
of isotopes of Li, Be, B.
Examples of using evaluated cross sections are given in [3,6]. 

\section{Results}
Figs.\ 1-3 show production cross sections of isotopes of Li, Be, B
for most important channels where, at least, several data points are available. 
Our evaluated cross sections are shown together with data
and results of frequently used semi-empirical systematics [10,14].
The individual cross section evaluations were tested against isobaric
(A = 7, 10) and cumulative data where available.
The low-energy part of production cross sections is often resonance-shaped,
which may be important in case of reacceleration models [11]; this is 
usually ignored in propagation calculations. Some data on
production of Li, Be, B in reactions with other nuclei can be found in
[8,9,12,13] and in compilation [2].

The authors are grateful to W.\ R.\ Webber for providing
results of his cross sections calculations. This work
was supported in part by a NASA Astrophysics Theory Program grant
and the US Department of Energy.

\begin{figure}[!tb]
  \begin{center}
    \includegraphics[width=0.49\textwidth]{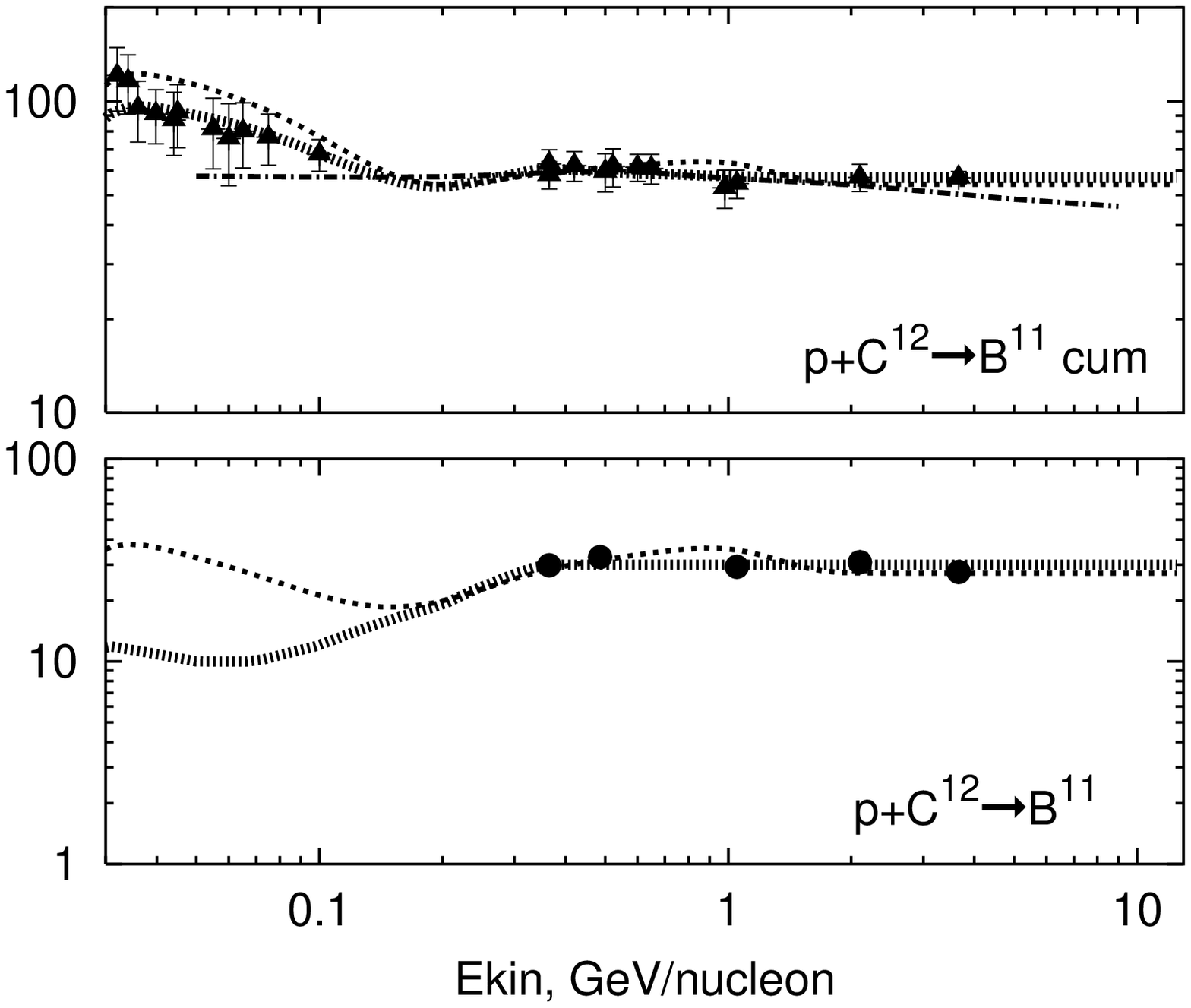}\hfill
    \includegraphics[width=0.49\textwidth]{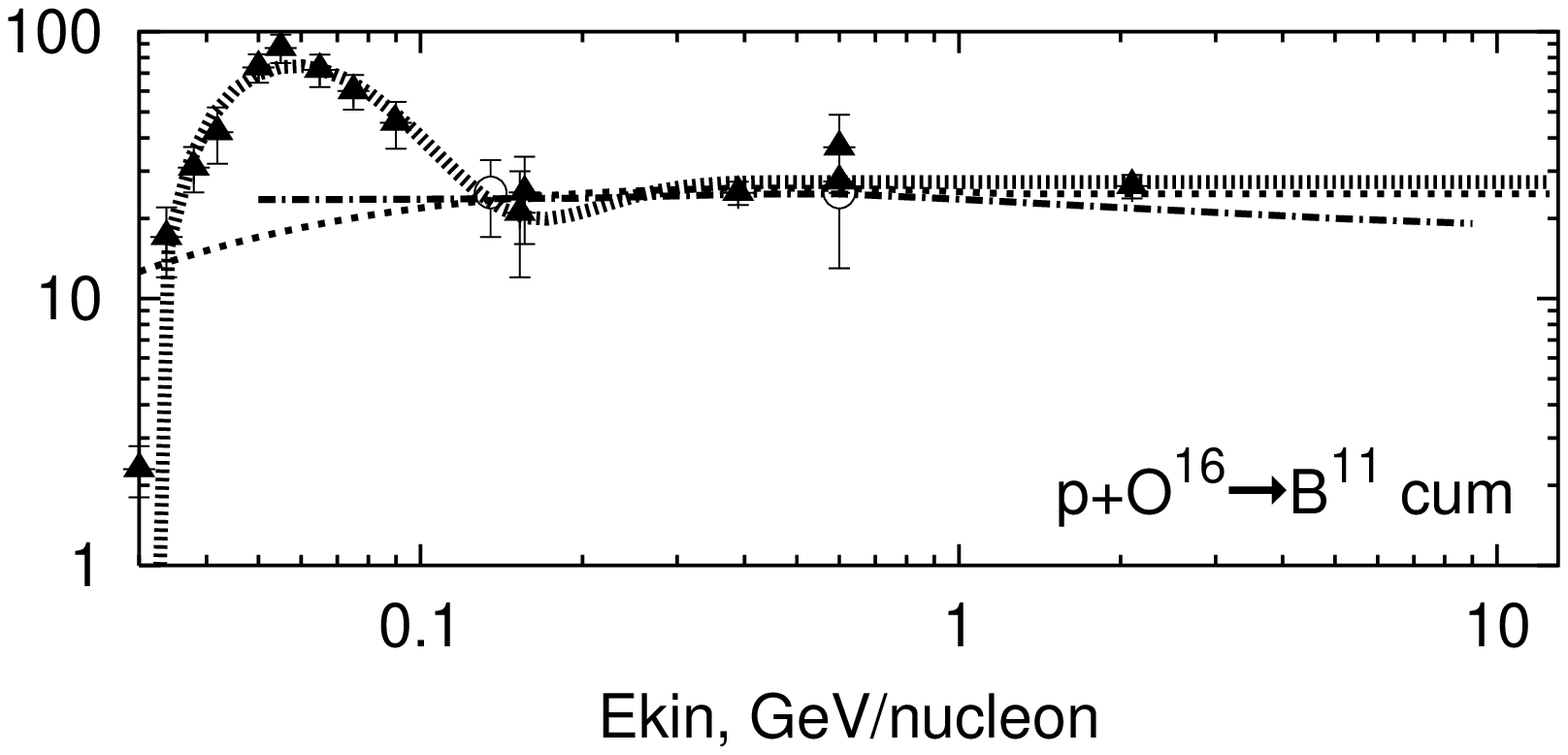}
  \end{center}
  \vspace{-0.5pc}
  \vspace{-2mm}
  \caption{The individual and cumulative production cross sections (mbarn) of B$^{11}$
vs.\ kinetic energy.
Lines: \lower-2pt\hbox{\tiny $\mid\mid\mid\mid\mid$} -- evaluated cross sections, 
\lower-2pt\hbox{\Large \bf ...} -- [10], 
{\Large -\lower-2pt\hbox{\bf .}-} -- [14]. Data: 
$\triangle$\hskip-0.7em$\bullet$\hskip-0.66em{\Large \bf .}\lower-4pt\hbox{\hskip-0.33em{\Large \bf .}}\hskip-0.31em{\Large \bf .}
-- isobaric cross section [9], 
\lower-1pt\hbox{\scriptsize $\bigcirc$} -- cross section on a natural
sample (compilation [2]),
\lower2pt\hbox{\LARGE $\bullet$} -- individual cross section [1,7-9,12,13].
}
\end{figure}

\begin{figure}[!tb]
  \begin{center}
    \includegraphics[width=0.49\textwidth]{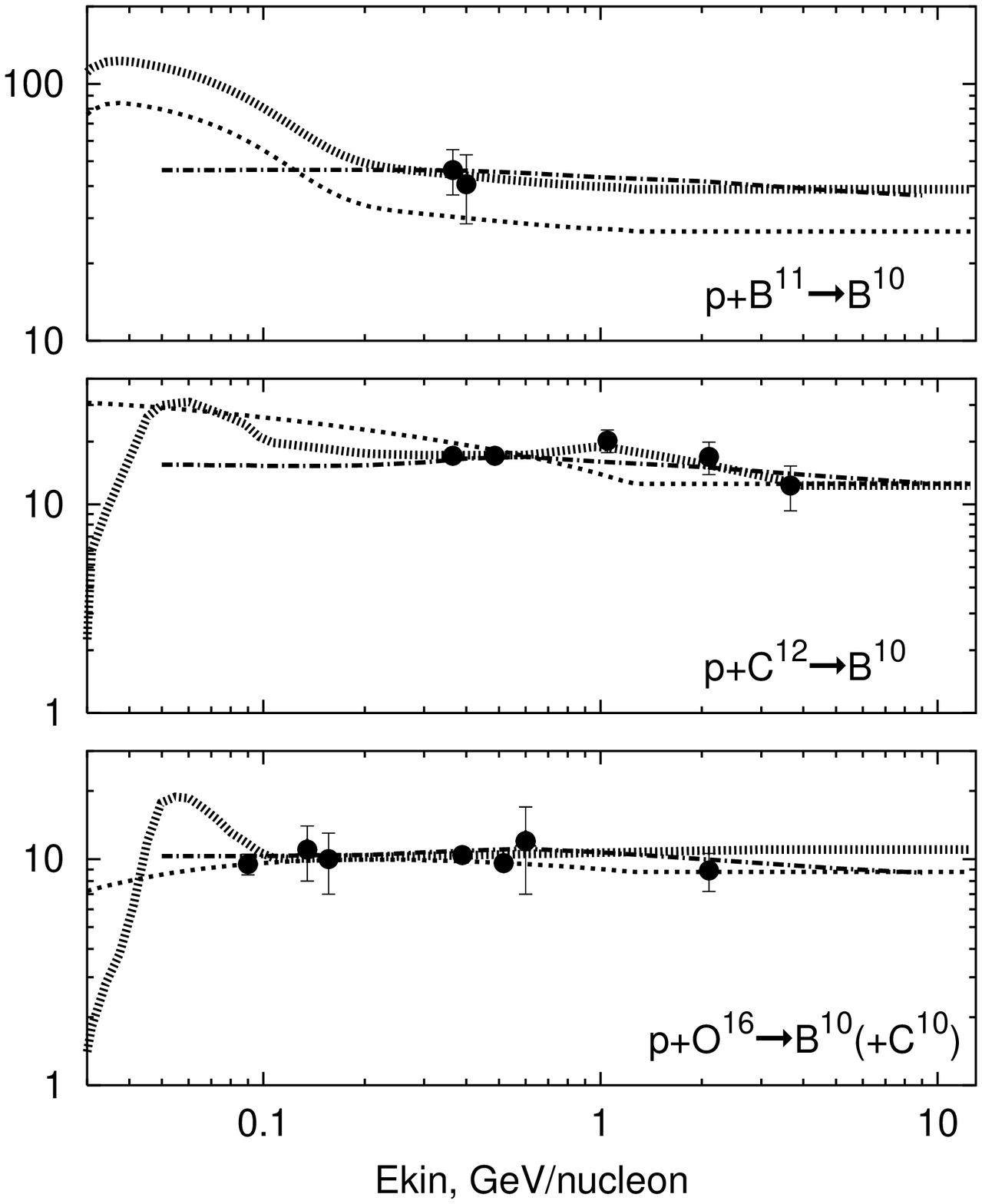}\hfill
    \includegraphics[width=0.49\textwidth]{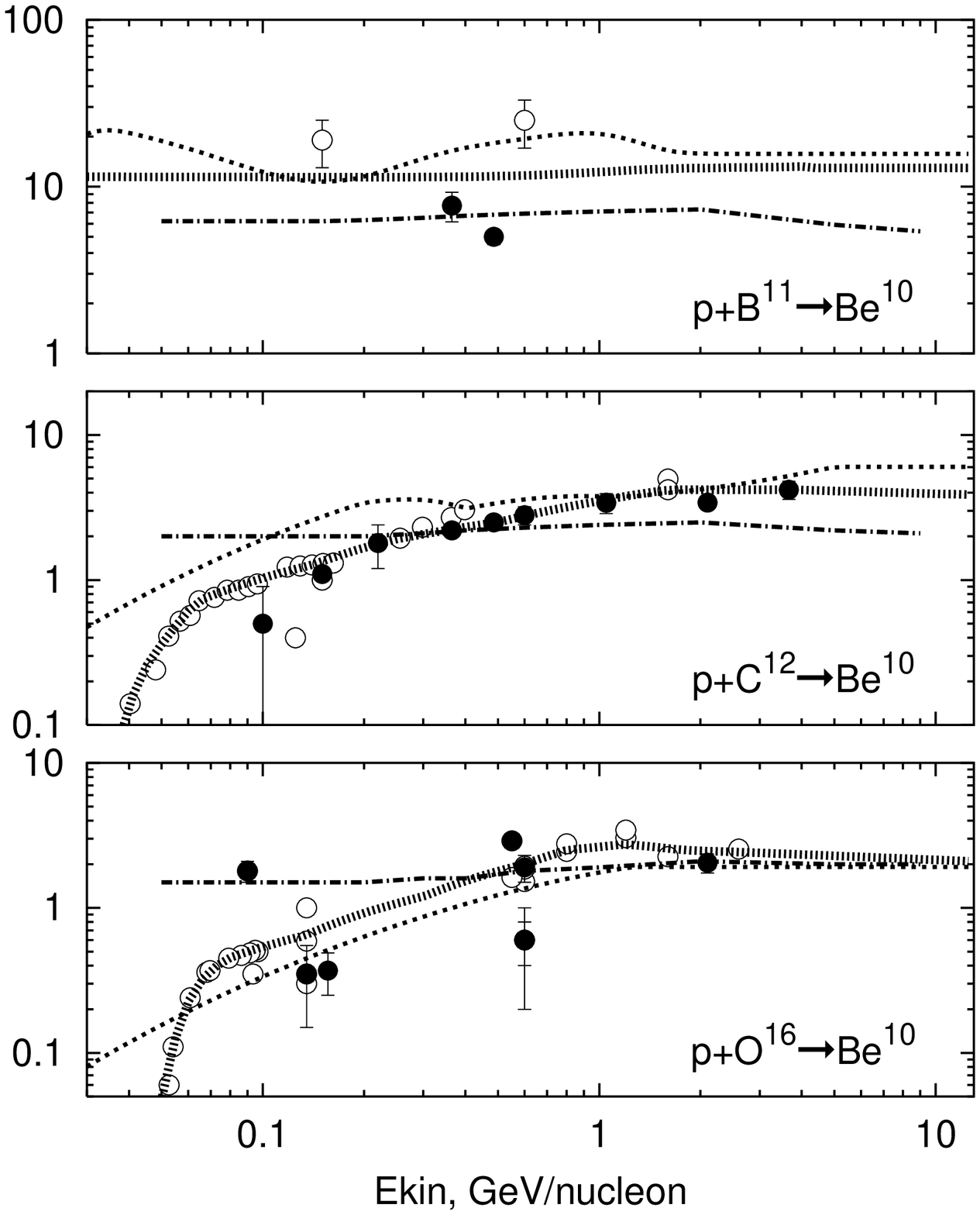}
    \includegraphics[width=0.49\textwidth]{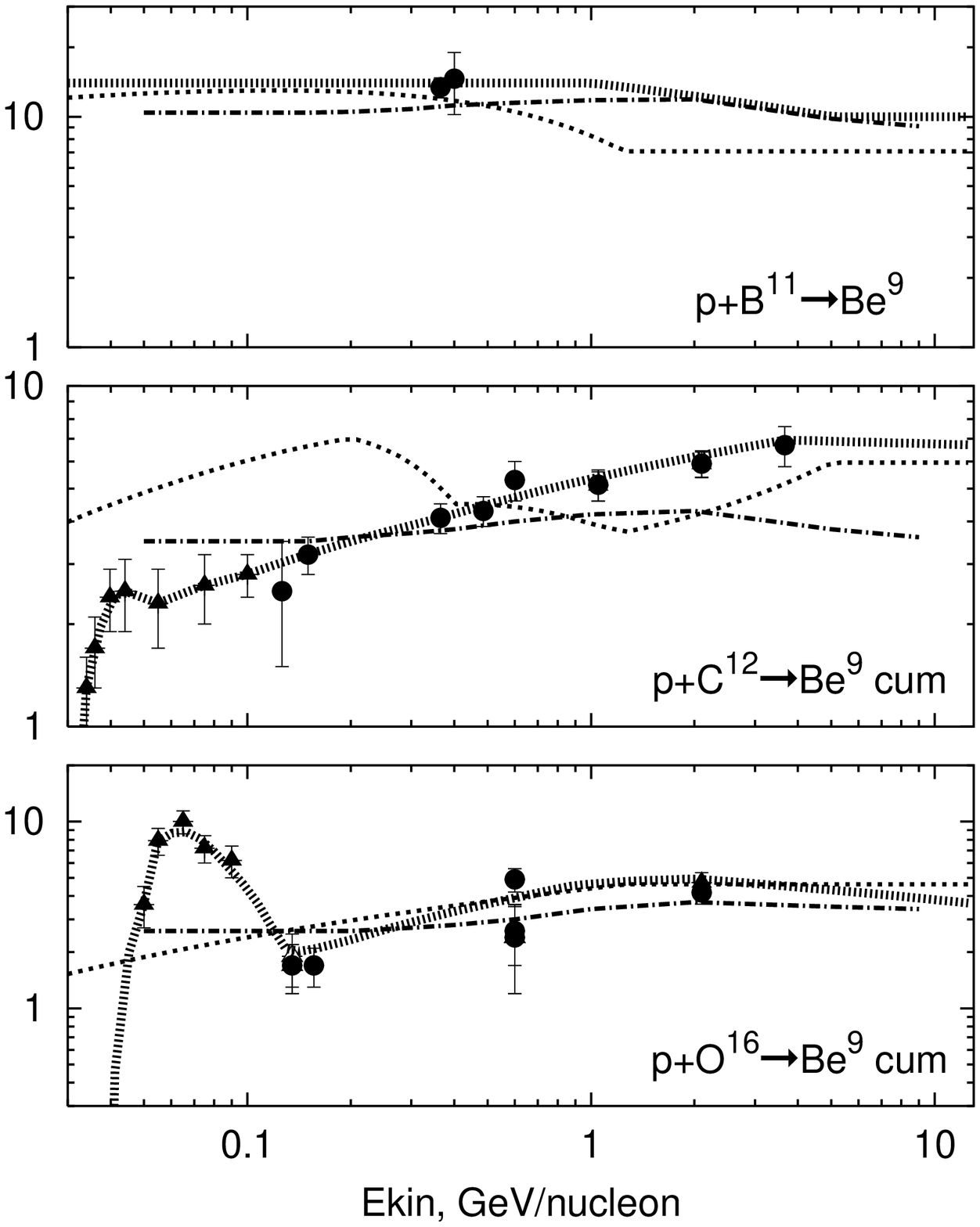}\hfill
    \includegraphics[width=0.49\textwidth]{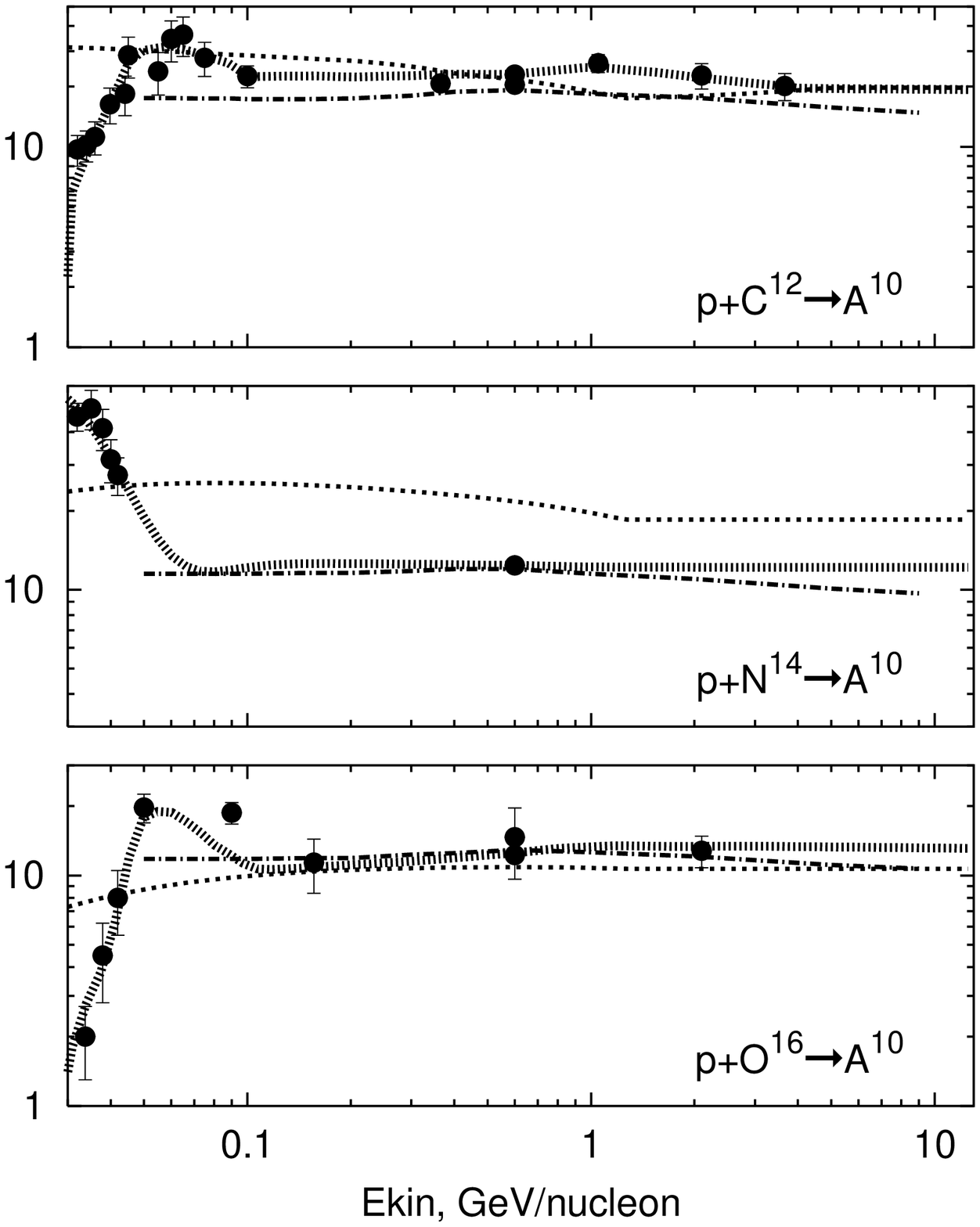}
  \end{center}
  \vspace{-0.5pc}
  \caption{The production cross sections (mbarn) of B$^{10}$,
Be$^{9,10}$, and isobaric cross section A$^{10}$=C$^{10}$+B$^{10}$+Be$^{10}$.
Lines and data symbols are coded as in Fig.~1.
}
\end{figure}

\begin{figure}[!tb]
  \begin{center}
    \includegraphics[width=0.49\textwidth]{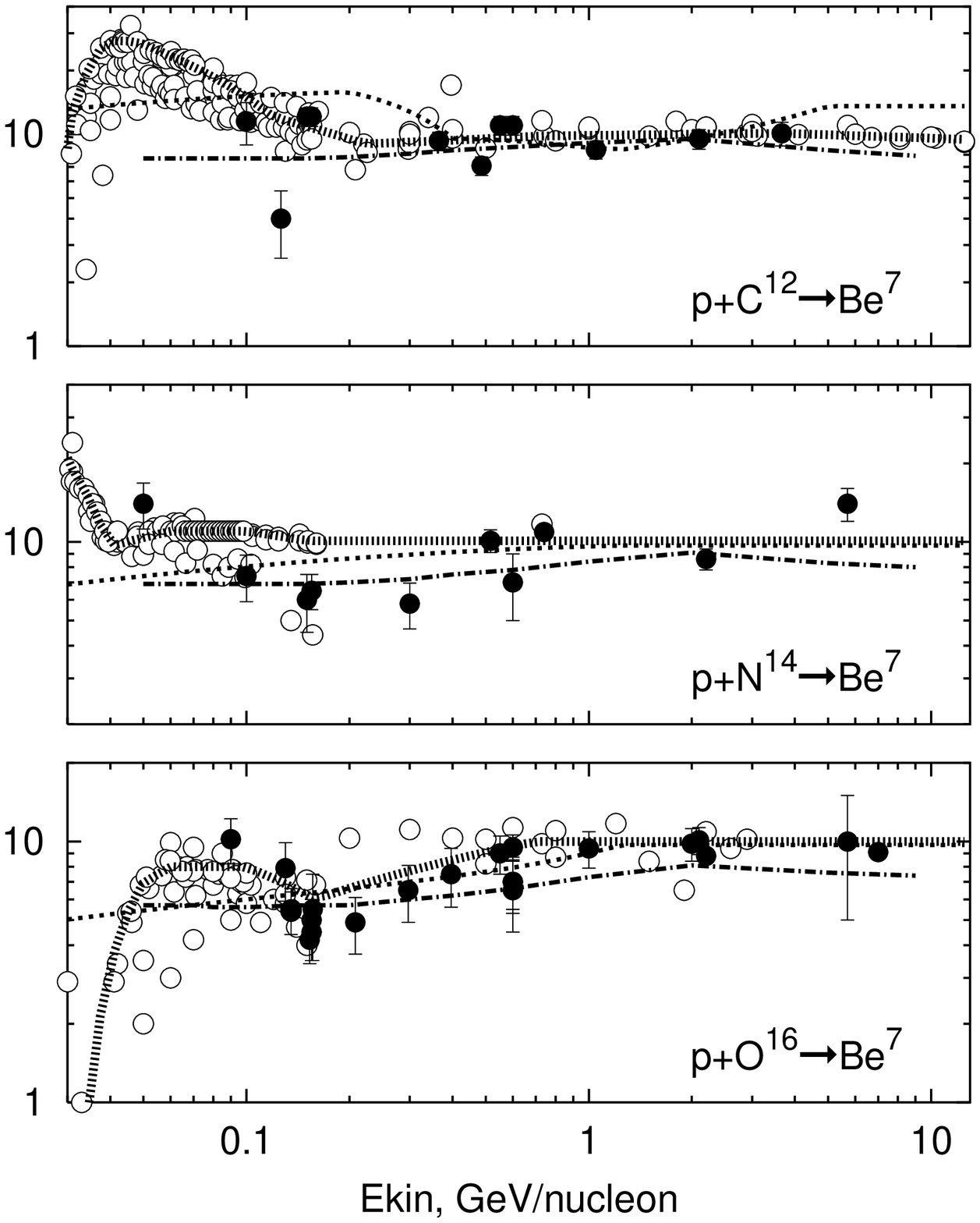}\hfill
    \includegraphics[width=0.49\textwidth]{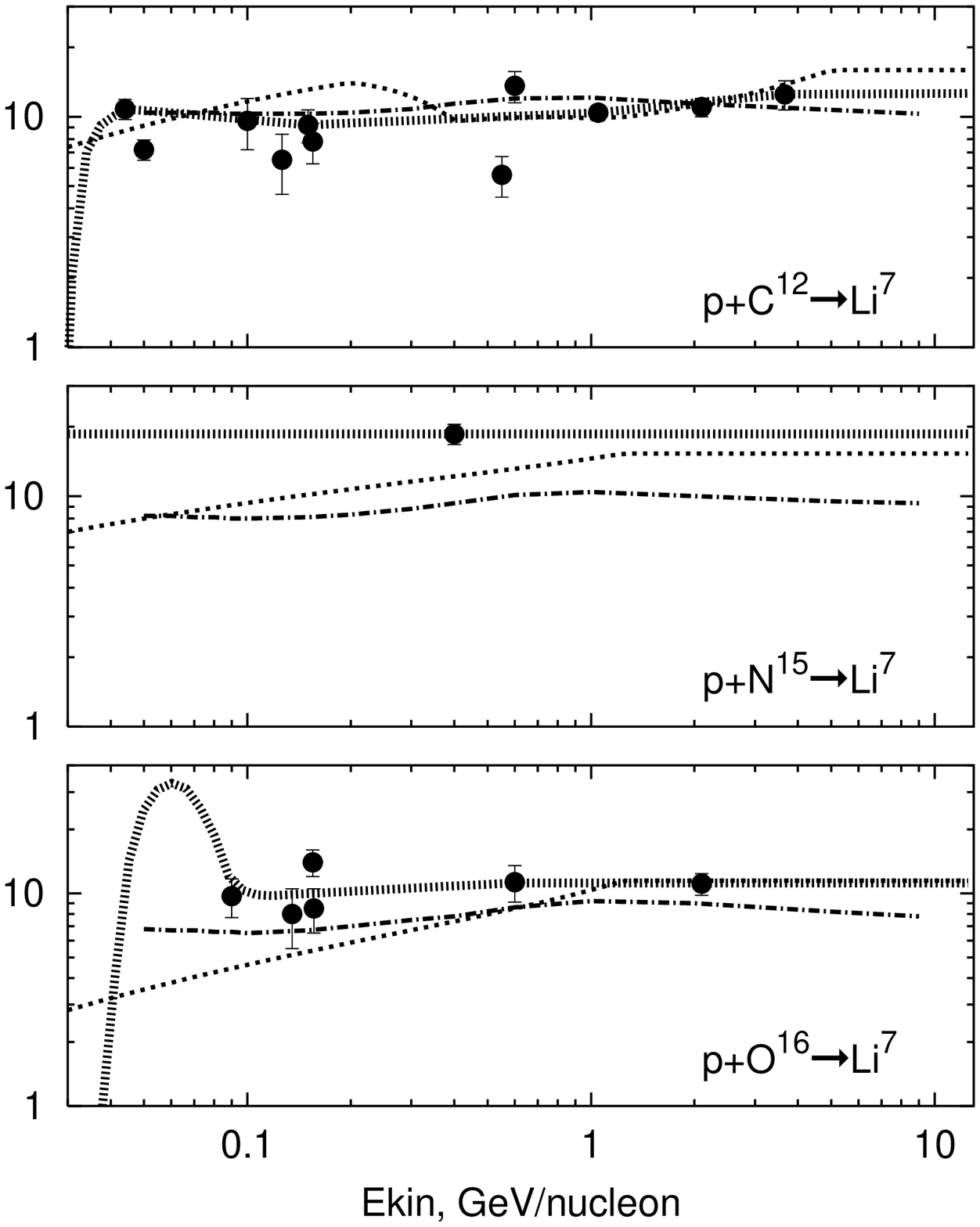}
    \includegraphics[width=0.49\textwidth]{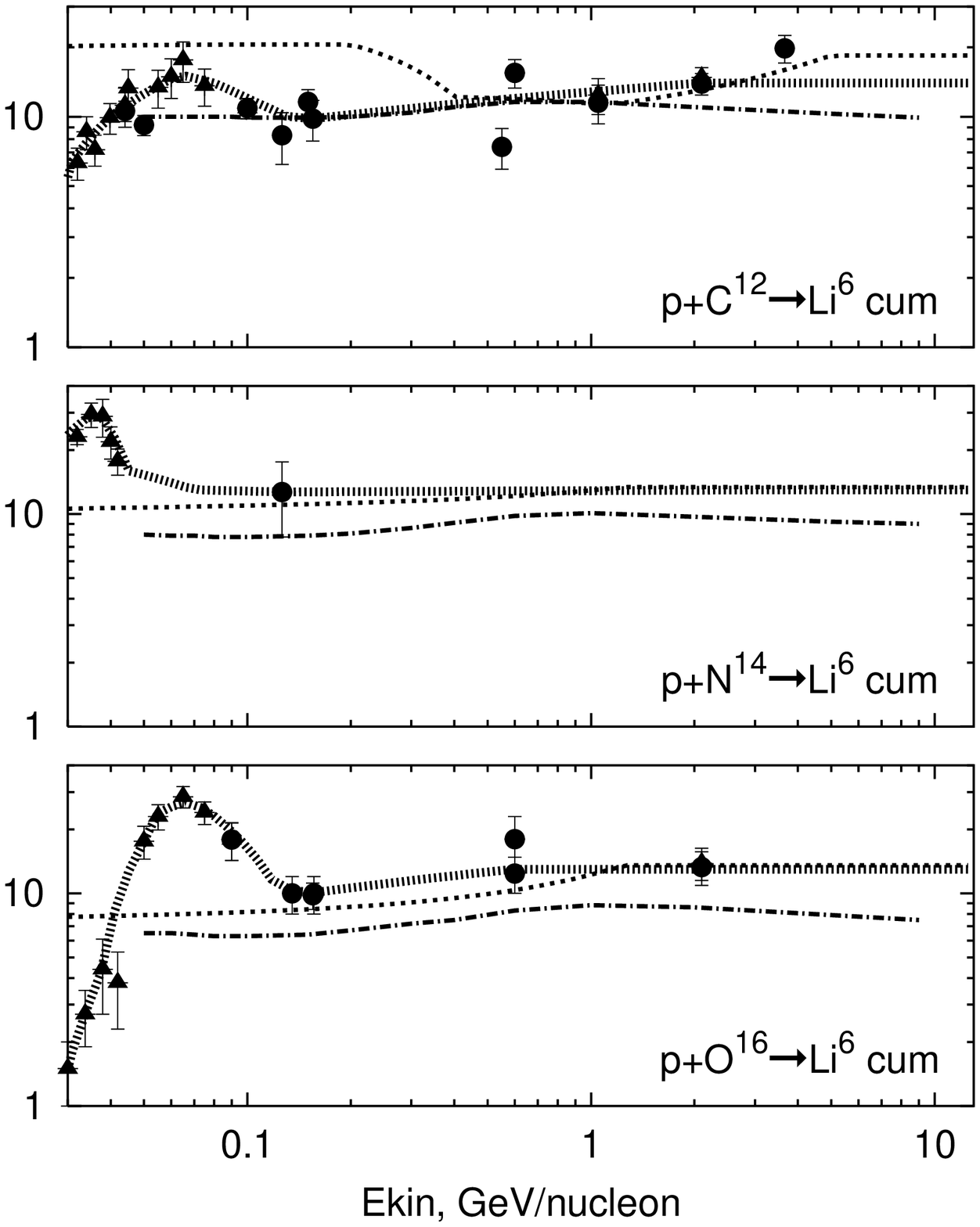}\hfill
    \includegraphics[width=0.49\textwidth]{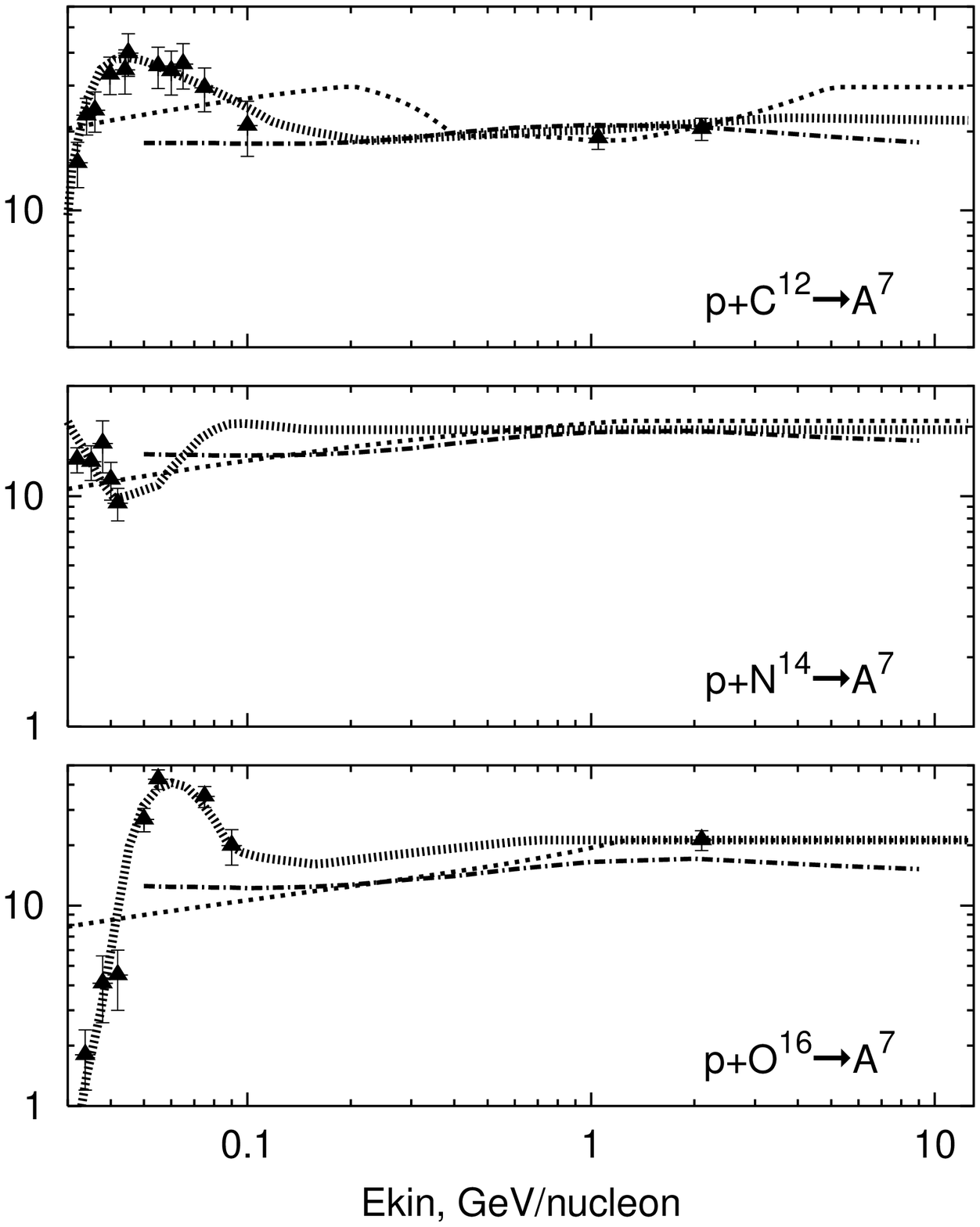}
  \end{center}
  \vspace{-0.5pc}
  \caption{The production cross sections (mbarn) of Be$^7$,
Li$^{6,7}$, and isobaric cross section A$^7$=Be$^7$+Li$^7$+He$^7$.
Lines and data symbols are coded as in Fig.~1.
}
\end{figure}

\section{References}
\re
1.\ Korejwo A.\ et al.\ 1999, in Proc.\ 26th ICRC (Salt Lake City), 4, 267
\re
2.\ Mashnik S.G., Sierk A.J., Van Riper K.A., Wilson W.B.\ 1998, 
in Proc.\ 4th Workshop on Simulated Accelerator Radiation Environments,
ed.\ Gabriel T.A.\ (ORNL: Oak Ridge, TN), 151
\re
3.\ Moskalenko I.V., Mashnik S.G., Strong A.W.\ 2001, 
in Proc.\ 27th ICRC (Hamburg), 1836 
\re
4.\ Moskalenko I.V., Strong A.W., Ormes J.F., Potgieter M.S.\ 2002, ApJ 565, 280
\re
5.\ Moskalenko I.V., Strong A.W., Mashnik S.G., Ormes J.F.\ 2003, ApJ 586, 1050
\re
6.\ Moskalenko I.V., Strong A.W., Mashnik S.G., Jones F.C.\ 2003, these Proc.
\re
7.\ Olson D.L.\ et al.\ 1983, Phys. Rev. C 28, 1602
\re
8.\ Radin J.R., Gradsztajn E., Smith A.R.\ 1979, Phys. Rev. C 20, 787
\re
9.\ Read S.M., Viola V.E.\ 1984, Atom.\ Data \& Nucl.\ Data Tables 31, 359
\re
10.\ Silberberg R., Tsao C.H., Barghouty A.F.\ 1998, ApJ 501, 911
\re
11.\ Strong A.W., Moskalenko I.V.\ 1998, ApJ 509, 212 
\re
12.\ Webber W.R., Kish J.C., Schrier D.A.\ 1990, Phys.\ Rev.\ C 41, 547
\re
13.\ Webber W.R.\ et al.\ 1998, ApJ 508, 949
\re
14.\ Webber W.R., Soutoul A., Kish J.C., Rockstroh J.M.\ 2003, ApJS 144, 153

\endofpaper
\end{document}